\documentclass[10pt, paper=a4, UKenglish]{article}
\usepackage{graphicx}
\usepackage{paralist}
\def\Title#1{\begin{center} {\Large #1 } \end{center}}
\def\Author#1{\begin{center}{ \sc #1} \end{center}}
\def\Address#1{\begin{center}{ \it #1} \end{center}}

\newcommand\pubblock{\rightline{\begin{tabular}{l} Proceedings of the CTD/WIT 2019\\ \pubnumber\\
         \pubdate  \end{tabular}}}

\newenvironment{Abstract}{\begin{quotation} \begin{center} 
             \large ABSTRACT \end{center}\bigskip 
      \begin{center}\begin{large}}{\end{large}\end{center} \end{quotation}}

\newenvironment{Presented}{\begin{quotation} \begin{center} 
             PRESENTED AT\end{center}\bigskip 
      \begin{center}\begin{large}}{\end{large}\end{center} \end{quotation}}

\def\beq{\begin{equation}}
\def\eeq#1{\label{#1}\end{equation}}
\def\eeqn{\end{equation}}

\def\beqa{\begin{eqnarray}}
\def\eeqa#1{\label{#1}\end{eqnarray}}
\def\eeqan{\end{eqnarray}}

\let\bar=\overbar

\def\Dslash{\not{\hbox{\kern-4pt $D$}}}
\def\dslash{\not{\hbox{\kern-2pt $\del$}}}

\def\msb{{\bar{\ssstyle M \kern -1pt S}}}

\usepackage{color}
\usepackage{lineno}
\usepackage{subfig}
\usepackage{hyperref}

\newcommand\pubnumber{PROC-CTD19-013}

\newcommand\pubdate{\today}

\def\affiliation{
on behalf of the ALICE Collaboration, \\
Istituto Nazionale di Fisica Nucleare, Sezione di Torino \\
Via P.Giuria 1, 10125, Torino, Italy; \\
European Organization for Nuclear Research (CERN), \\
1211 Geneva 23, Switzerland}

\newcommand{\conference}{Connecting the Dots and Workshop on Intelligent Trackers (CTD/WIT 2019)\\
Instituto de F\'isica Corpuscular (IFIC), Valencia, Spain\\ 
April 2-5, 2019}

\usepackage{fancyhdr}
\definecolor{mygrey}{RGB}{105,105,105}
\begin{document}


\large
\begin{titlepage}
\pubblock

\vfill
\Title{Global Track Reconstruction and Data Compression Strategy in ALICE for LHC Run 3}
\vfill

\Author{David Rohr}
\Address{\affiliation}
\vfill

\begin{Abstract}
In LHC Run 3, ALICE will increase the data taking rate significantly, from an approximately 1 kHz trigger readout in minimum-bias Pb--Pb collisions to a 50 kHz continuous readout rate.
The reconstruction strategy of the online-offline computing upgrade foresees a synchronous online reconstruction stage during data taking, which generates the detector calibration, and a posterior calibrated asynchronous reconstruction stage.
The huge amount of data requires a significant compression in order to store all recorded events.
The aim is a factor 20 compression of the TPC data, which is one of the main challenges during synchronous reconstruction.
In addition, the reconstruction will run online, processing 50 times more collisions than at present, yielding results comparable to current offline reconstruction.
These requirements pose new challenges for the tracking, including the continuous TPC readout, more overlapping collisions, no a priori knowledge of the primary vertex position and of location-dependent calibration during the synchronous phase, identification of low-momentum looping tracks, and a distorted refit to improve track model entropy coding.
At the 2018 workshop, the TPC tracking for Run 3 was presented, which matches the physics performance of the Run 2 offline tracking.
It leverages the potential of hardware accelerators via the OpenCL and CUDA APIs in a shared source code for CPUs and GPUs for both reconstruction stages.
Porting more reconstruction steps like the remainder of the TPC reconstruction and tracking for other detectors to GPU will shift the computing balance from traditional processors towards GPUs.
These proceedings focus on the global tracking strategy, including the ITS and TRD detectors, offloading more reconstruction steps onto GPU, and the approaches taken to achieve the necessary data compression.
\end{Abstract}

\vfill

\begin{Presented}
\conference
\end{Presented}
\vfill
\end{titlepage}
\def\thefootnote{\fnsymbol{footnote}}
\setcounter{footnote}{0}
%

\normalsize 

\section{Introduction}
\label{intro}

ALICE (A Large Ion Collider Experiment) \cite{bib:alice} is one of the four main experiments at the LHC (Large Hadron Collider) at CERN.
It is dedicated to study heavy ion collisions at unprecedented energies.
Currently, the LHC is in the second long shutdown phase (LS2) and will restart operation in 2021 with an increased Pb--Pb interaction rate.
Meanwhile, ALICE is upgrading many of its main detectors and computing systems~\cite{bib:aliceupgrade}.
In particular, the TPC (Time Projection Chamber) will switch from MWPC (Multi Write Proportional Chambers) to a GEM (Gas Electron Multiplier) readout~\cite{bib:tpcrun3tdr}, which will enable continuous readout at 50 kHz Pb--Pb interaction rate.
The ITS (Inner Tracking System) will be replaced by 7 layers of silicon MAPS (Monolithic Active Pixel Sensor) detectors.
Additionally,  the online computing infrastructure and the whole computing scheme will be changed to cope with the increased data rates~\cite{bib:o2tdr}.

In LHC Run 3, ALICE will record minimum-bias Pb--Pb collisions at 50 kHz in continuous read-out, as compared to the Run 2 trigger read-out rate that was below 1 kHz.
The design foresees the storage of all collisions without any software trigger, in order to have full access to rare physics signals.
The continuous read-out of pp collisions will happen at rates between 200 kHz and 1 MHz.
Sophisticated data compression in the online computing farm is required in order to stay within the allotted storage capacity.
In contrast to the simpler data compression ALICE applied in Run 2, the new compression scheme relies on the event reconstruction.
This makes online reconstruction and to some extent online calibration inevitable.
Overall, this creates a large computational challenge of processing, in real time, 50 to 100 times more events per second than in Run 2, thus demanding that the reconstruction and compression algorithms must be more complex than in the HLT (High Level Trigger) \cite{bib:hltpaper} during Run 2.

A new online computing farm will be installed, comprising roughly 750 computing servers and 1500 GPUs (Graphics Processing Units).
The new computing paradigm foresees the utilization of this farm for synchronous event processing during data taking.
The synchronous processing performs the data compression, stores all compressed data to a disk buffer, and aggregates all required quantities for the calibration.
The latter makes the offline calibration pass over the data, which was executed during Run 2, obsolete.
It follows a post-processing step to create the final detector calibration.
During times without beam, e.\,g.~during LHC turn-around, technical stop, etc., or when the synchronous processing does not require the full farm, like during pp data taking, the available computing resources will be used for the asynchronous reprocessing of the data from the disk buffer yielding the final reconstruction output.
Synchronous and asynchronous stages will employ the same software and algorithms, however with different settings, final calibration, and additional reconstruction steps.

\begin{figure}[!b]
  \centering
  \subfloat[]{\includegraphics[width=0.47\linewidth]{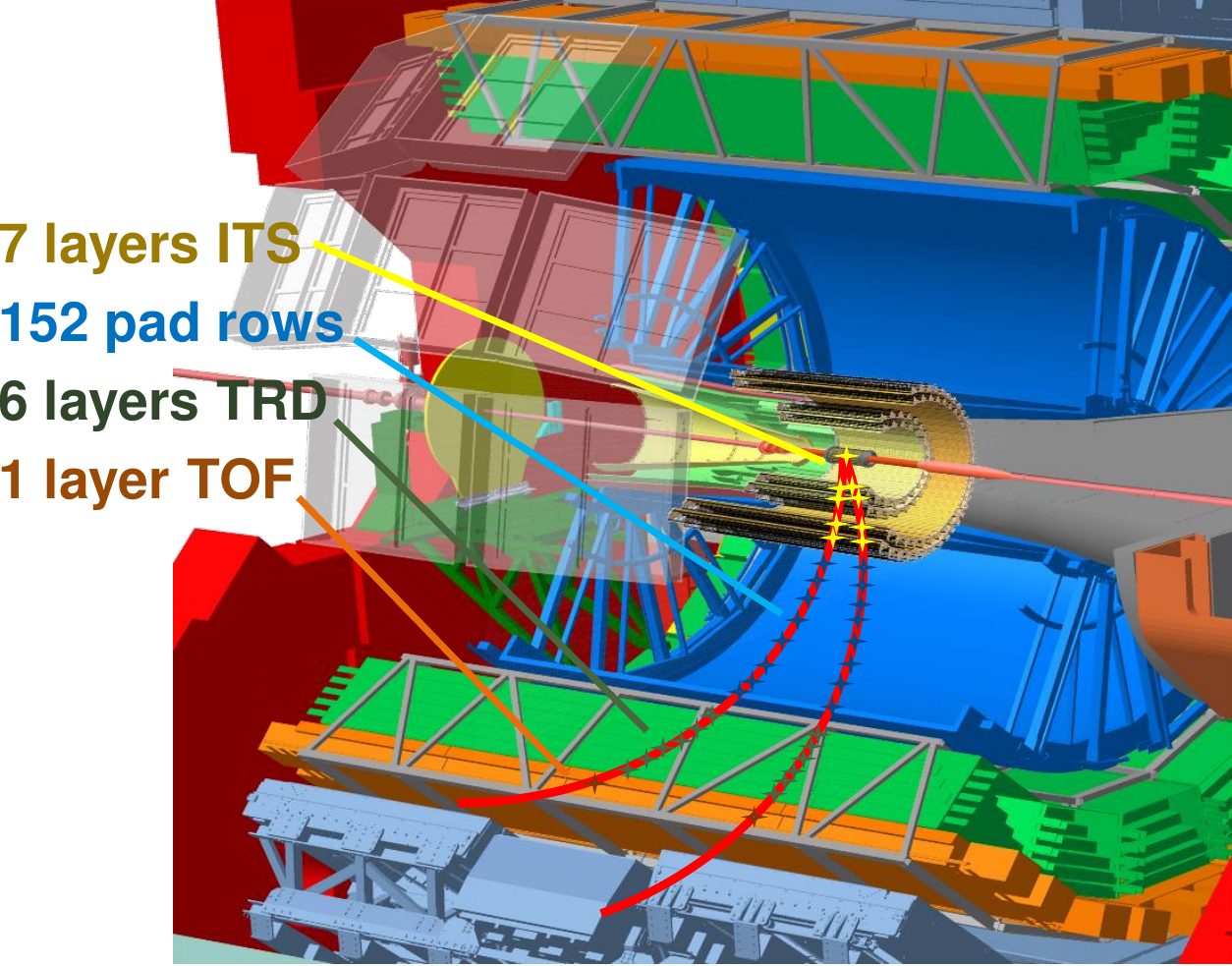}}
  \qquad
  \subfloat[]{\includegraphics[width=0.47\linewidth]{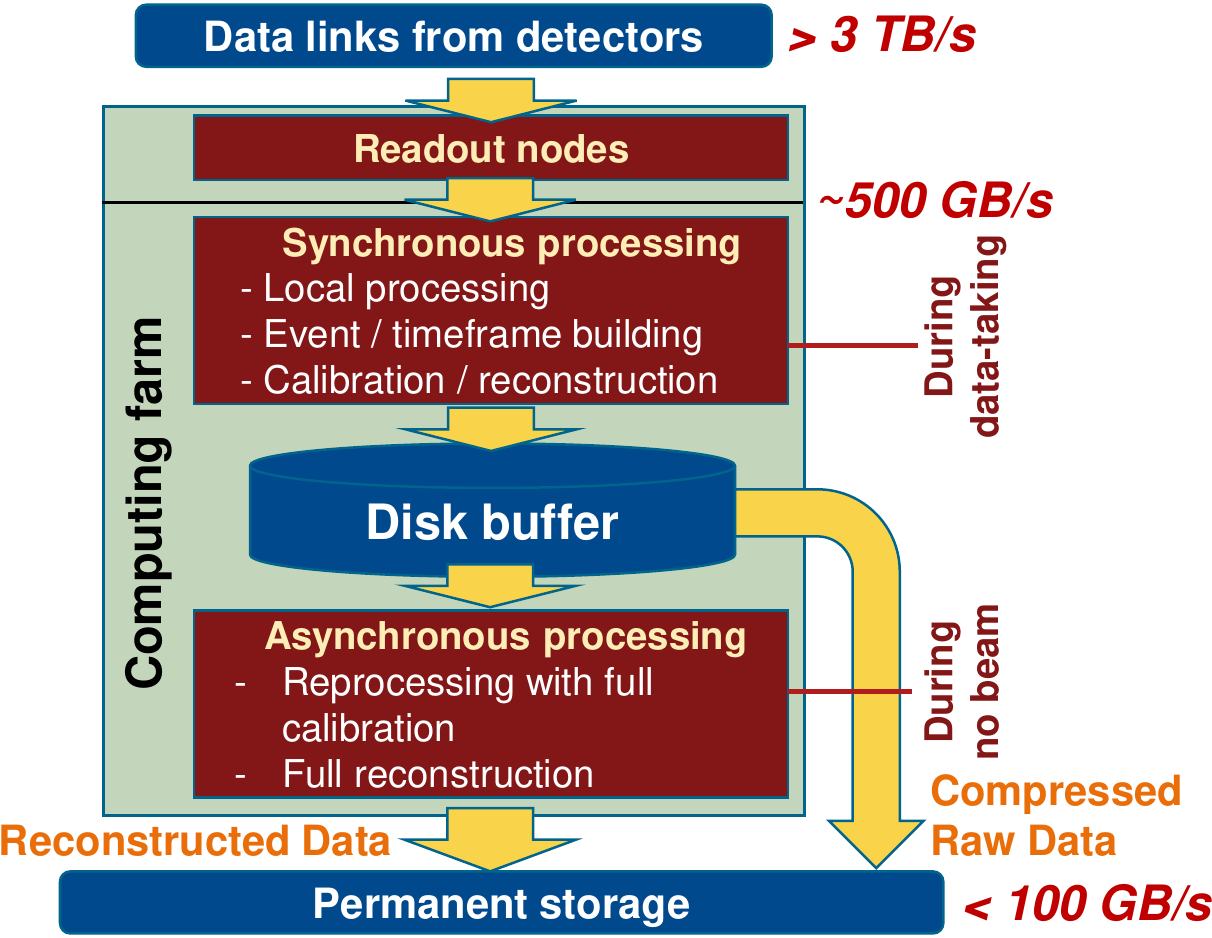}}
  \caption{(a) Tracking detectors of ALICE, (b) Computing scheme of ALICE computing upgrade.}
  \label{fig:overview}
\end{figure}

Figure~\ref{fig:overview} gives an overview of the tracking detectors of ALICE in the central-barrel region, as well as of the online computing scheme.
In addition to the TPC and the ITS, the TRD (Transition Radiation Detector) and TOF (Time Of Flight) are used to calibrate the TPC~\cite{bib:lhcp2017} and for better momentum resolution.
The track reconstruction and data compression will be the most time-consuming steps of the online reconstruction.
ALICE will employ GPUs to speed up the tracking, which will provide significant cost-savings compared to a traditional approach with CPUs.

\section{Reconstruction Strategy}

The synchronous processing will serve the purposes of calibration and data compression, where the largest contributor to the data volume by far is the TPC.
Since the TPC data reduction will be based on reconstructed tracks, the synchronous processing must perform full TPC tracking.
In addition, the calibration is based on refitting tracks in the ITS, TPC, TRD, and TOF~\cite{bib:lhcp2017}, and thus requires tracking in all four detectors.
However, a very limited subset of the events, in the order of 1\%, will be enough to gain sufficient statistics.
Therefore, the main synchronous workload will be TPC tracking.
ALICE must be able to cope with the maximum data rate in real time which will be 50 kHz ob Pb-Pb interactions during the beginning of a heavy ion LHC fill.
The online computing farm will be dimensioned accordingly, offering a sufficient number of GPUs for the required computing power.

During the asynchronous processing, the TPC tracking will not be dominant.
Due to the removal of hits in the TPC not used for physics analysis, as described in Sec.~\ref{sec:compression}, the TPC tracking will be faster than during the synchronous stage.
In contrast, the full ITS tracking is required, which is computationally expensive because of the more complicated combinatorics in the inner region of the detector.
Consequently, the ITS tracking will also utilize GPUs~\cite{bib:itsgpu}.

The TPC tracking is derived from the current Run 2 HLT TPC tracking on GPUs~\cite{bib:hltpaper} and has been adopted and improved for the Run 3 setting~\cite{bib:ctd2018,bib:chep2018}.
In addition, the Run 2 HLT TPC d$E$/d$x$ algorithm has been adapted to run on GPUs.
The reconstruction steps described so far, TPC tracking, ITS tracking, TPC compression, and d$E$/d$x$ calculation, are the \textbf{baseline scenario} for the GPU reconstruction.
Offloading these steps to the graphics card will ensure that ALICE has sufficient computing resources to cope with Run 3 data rates.

\begin{figure}[!h]
  \centering
  \includegraphics[width=0.8\textwidth]{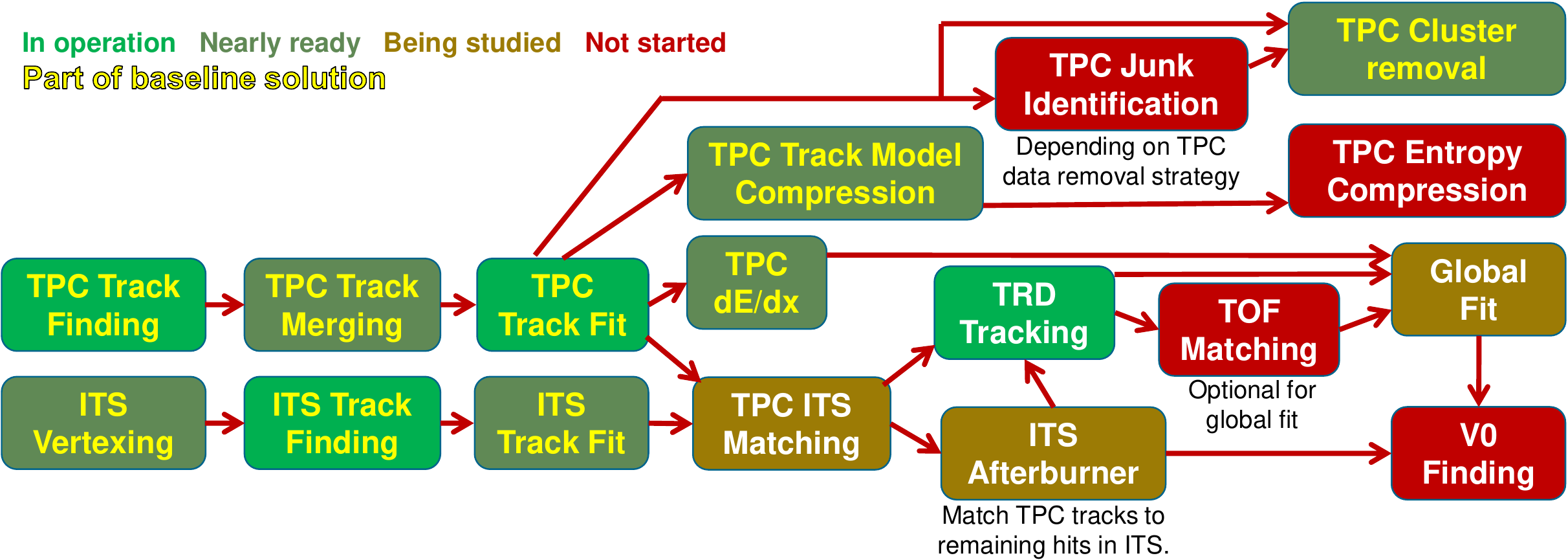}
  \caption{Reconstruction steps of global tracking and compression chain with significant computing time.}
  \label{fig:chain}
\end{figure}

\looseness=-1
The number of GPUs is chosen to match the computing demands of the synchronous phase.
However, the baseline solution will probably not fully utilize the GPUs during the asynchronous phase, since the TPC tracking will be faster than during the synchronous phase, and many reconstruction steps will remain on the CPU.
Therefore, a \textbf{full scenario}, where as many reconstruction steps as possible use the GPU, given that they are available in the online computing farm, is also considered.
A promising candidate is the full tracking chain, which includes track matching, TRD and TOF reconstruction, secondary vertexing, and the global fit.
The majority of the data will already be present on the GPU and with all of the tasks being inherently parallel, they are well suited for execution on a GPU.
This would allow for offloading the full tracking chain to the GPU in one go and processing it there without the need for intermediate data transfer forth and back.
Figure~\ref{fig:chain} shows the reconstruction steps related to tracking and the current state of their GPU offloading efforts.

In general, the offloading of the entire chain is not needed.
The required GPU algorithms are implemented step-by-step from the beginning of the chain to the end.
For example, it is no showstopper  if the secondary vertexing or the global fit at the end of the chain do not run on the GPU.
All previous steps would run on the GPU and the data would be shipped back to the host, not at the very end of the chain but few steps before.
Besides the tracking, reconstruction for other detectors could potentially benefit from GPUs, but our main focus is tracking, with other developments happenening independently.

\section{TPC GPU Tracking}

\begin{figure}[htb]
  \centering
  \includegraphics[width=1.0\textwidth]{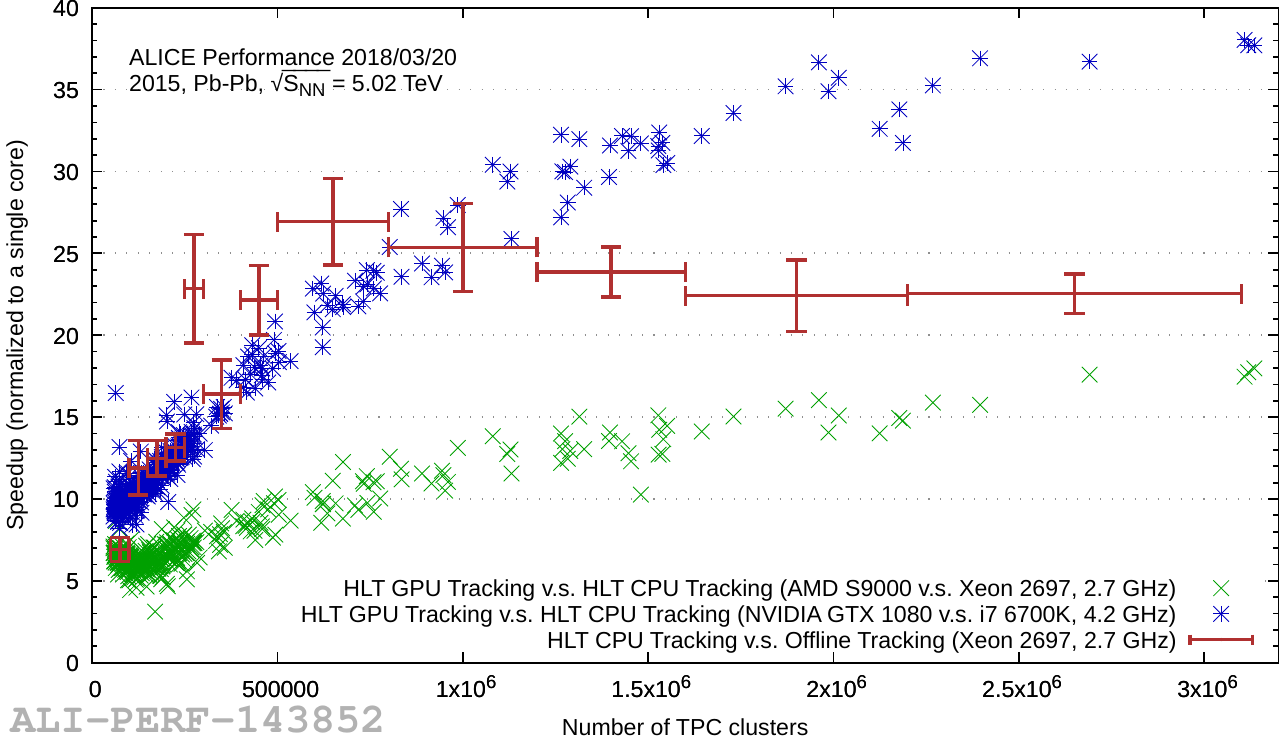}
  \caption{Speedup of the ALICE TPC track finding on GPU normalized to a single CPU core.}
  \label{fig:speedup}
\end{figure}

The TPC tracking algorithm is derived from that of the Run 2 HLT~\cite{bib:hltpaper}.
Critical aspects of the latter were the independence from sole-vendors, avoidance of diverged code bases for CPU and GPU, the consistency of results on CPU and GPU, and the calibration.
Figure~\ref{fig:speedup} shows the speedup of the TPC track finding on a GPU normalized to a single CPU core~\cite{bib:chep2018}.

Since the beginning, the GPU tracking algorithm was implemented as generic C++ code, containing several macros which resolve to the GPU specific keywords.
Small wrappers for the CPU version and different GPU APIs like CUDA and OpenCL exist, which contain the implementation of the macros and the management code for initialization, data transfer, etc.~\cite{bib:generic}.
This ensured vendor-independence.

The first version showed small differences between the CPU and GPU version, originating from the concurrent processing.
These only affected technical aspects but not the physical results.
An example is a TPC hit in the vicinity of two TPC tracks being assigned to either one or the other depending on the order of processing.
These effects are mitigated as much as possible~\cite{bib:cnna}, but due to non-associative floating point arithmetic in combination with the \textit{-ffast-math} compiler flags, minor rounding differences are unavoidable.

While the final calibration will only be available for asynchronous reconstruction, a preliminary online calibration is required in the synchronous phase.
This has been tested for the TPC drift velocity calibration in Run 2~\cite{bib:hltpaper}.
The drift velocity, which is stable for a duration of about 15 minutes, is calibrated online over few minutes, and the obtained velocity is then used in the reconstruction for the following minutes.
At the very beginning of a data taking run, no calibration is available, and it must be estimated from previous runs.

\section{TPC Data Reduction}

\label{sec:compression}

The reduction of the TPC data size can be seen as several consecutive logical steps, although some are jointly implemented as one processing step:
\begin{compactenum}
 \item The TPC cluster finder converts the raw TPC ADC (Analog Digital Converter) values into hits, which are later used for the tracking.
  This step does not reduce the data size, but it is required for the steps thereafter.
 \item The cluster properties are converted from the single-precision floating point format, which is used in the reconstruction, to an integer format with exactly as many bits as needed for the intrinsic TPC resolution.
   Cluster size and charge are encoded with a dynamic precision with respect to their absolute value.
   All bits in the integer representation that lie $n$ bits after the first non-zero bit are truncated to zero, respecting proper rounding, with $n = 3$ for the size and $n = 4$ for the charge.
  \item The entropy is reduced in preparation for step 4.
   Clusters that are not assigned to tracks are ordered in $y$ and $z$ coordinates, and only the difference of the cluster position with respect to the previous cluster is stored.
   In high occupancy situations, like Pb-Pb collisions, the distribution of the differences strongly favors small values in contrast to the equally distributed absolute values, yielding a lower entropy.
   Clusters assigned to tracks are stored with the so-called track model compression, which does store the residuals of the cluster position to the extrapolated track instead of the absolute cluster positions~\cite{bib:lhcp2017}.
   These residuals contain even less entropy than the differences.
   Cluster charges can be stored with respect to the d$E$/d$x$ of the track and cluster sizes with respect to the average sizes of the clusters of a track.
 \item An entropy encoder compresses the entropy-reduced clusters..
   In Run 2, the HLT employed Huffman encoding.
   For Run 3, ALICE will switch to ANS encoding, which yields a roghly 5\% better compression for this use case.
 \item Clusters not used for physics analysis are discarded.
\end{compactenum}
\vspace{10pt}

Note that two of the steps require the reconstructed TPC tracks: the track-model and the removal of clusters not used for physics analysis.
The following types of clusters are candidates for removal:
\begin{compactenum}
  \item All clusters of tracks with $p_{\mathrm{T}} \leq 50$ MeV/$c$.
  \item Clusters of secondary legs of looping tracks with a $p_{\mathrm{T}}$ up to around 200 MeV/$c$.
  \item Clusters of track segments with a high inclination angle $\vert \phi \vert  \geq 70^{\circ}$ that are not used in the track fit.
  \item Clusters from noisy TPC pads.
  Permanently noisy pads are already masked at the read-out level, but this is not possible for pads which are temporarily noisy.
  \item Clusters from charge clouds produced by low-momentum protons (theoretically this is a subset of category 1, but the signature looks different).
\end{compactenum}
\vspace{10pt}

The cluster removal is executed in multiple steps.
The first step is the track reconstruction.
In order to be able to identify also the low-momentum looping tracks, the tracker was improved to reach a $p_{\mathrm{T}}$ of 10 MeV/$c$~\cite{bib:ctd2018}, although with decreasing efficiency below 20 MeV/$c$.
This produces the cluster to track attachment, which is resolved to the higher-momentum track if ambiguous.
Clusters within a tube of 1.5 cm around a track are called adjacent, where ambiguity is again resolved to higher-momentum tracks.
All clusters adjacent to the primary leg of a track with $p_{\mathrm{T}} > 50$ MeV/$c$ and a local inclination angle of $\vert \phi \vert  < 70^{\circ}$ are protected and will not be removed.
In case of looping tracks with multiple legs, only the primary, first leg is protected, but not the secondary legs.
Afterwards, there are two strategies:
\begin{compactitem}
  \item \textbf{Strategy B} is the most aggressive and will remove all remaining clusters.
  \item \textbf{Strategy A} will remove all non-protected clusters that are attached or adjacent to a track with $p_{\mathrm{T}} \leq 50$ MeV/$c$, a secondary leg of any track, or a track with high inclination angle of $\vert \phi \vert  \geq 70^{\circ}$.
    Since the tracking has a lower limit in $p_{\mathrm{T}}$, and since by design it can only identify tracks but not what is considered junk, posterior clean-up steps will run on the remaining unprotected clusters.
    A Hough-transform based approach is being investigated to identify the very low $p_{\mathrm{T}}$ tracks below 15 MeV/$c$.
    Additional heuristics will be needed to identify noise from TPC pads with a shifted baseline and charge clouds from low-momentum protons.
\end{compactitem}

While we aim for strategy A as it is the safer solution, strategy B will yield the better data reduction factor by construction.
In principle, since the tracking in the synchronous and the asynchronous phase run the same algorithm, the asynchronous phase should not be able to recover tracks (or prolong tracks) that were missed in the synchronous phase.
From this perspective, there is no sense in storing the additional clusters like in strategy A.
However, this is only true as long as the calibration is correct, and as long as there are no problems with the tracking algorithm.
By design, in strategy B, all tracks not found in the synchronous phase can never be recovered.
In contrast, strategy A is safe since it only rejects clusters that have been positively identified as not useful for physics analysis.
Since strategy B is basically a subset of strategy A, we go forward to implement strategy A, but in case we need additional data reduction and we are sufficiently confident in the maturity of the synchronous tracking, there is the option to switch to strategy B.

\begin{figure}[!htb]
  \centering
  \includegraphics[width=0.85\textwidth]{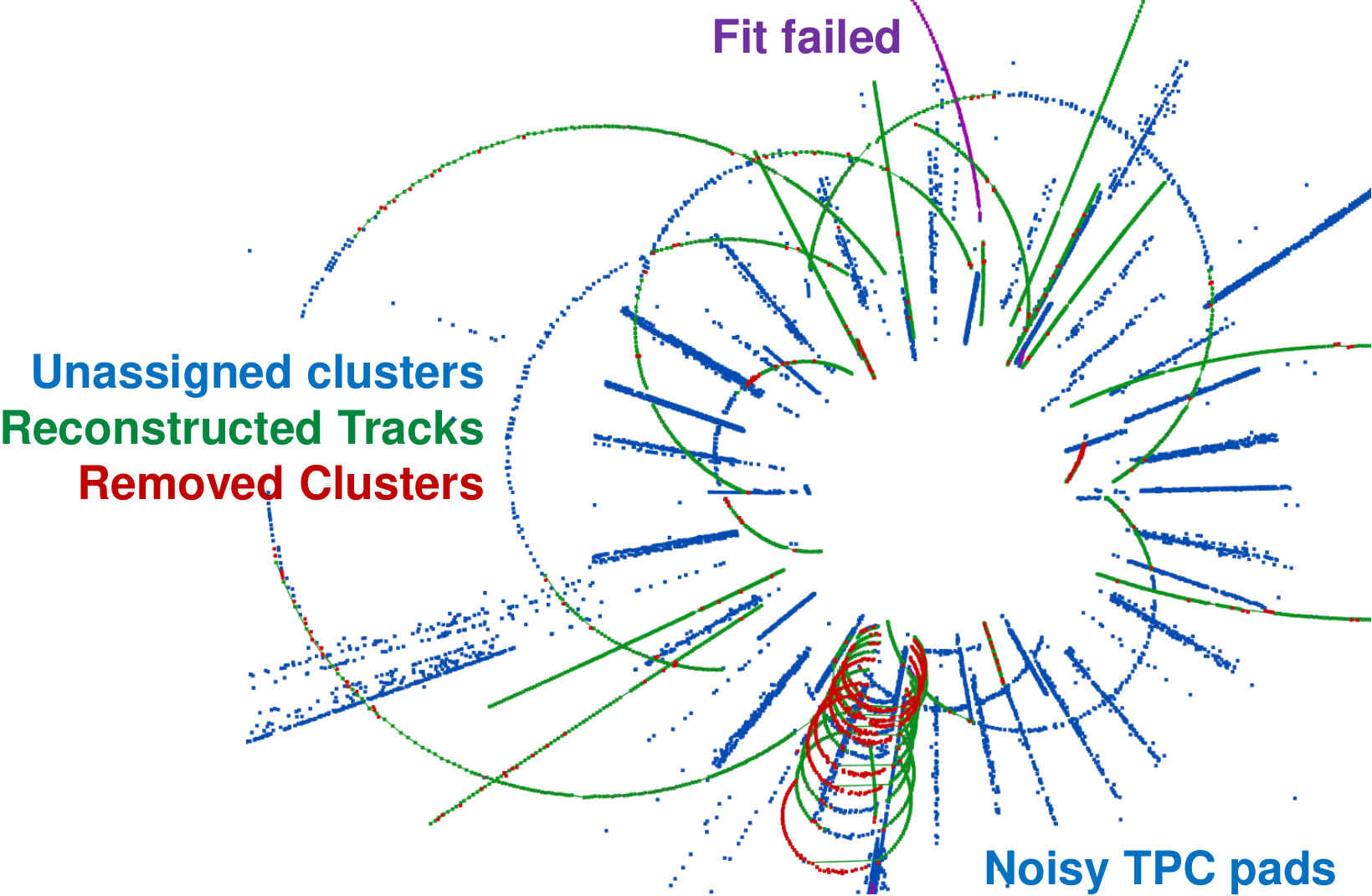}
  \caption{Example of low-$p_{\mathrm{T}}$ tracking and cluster rejection in single pp collision.}
  \label{fig:ex1}
\end{figure}

Irrespective of the strategy, efficient tracking down to low $p_{\mathrm{T}}$ is required.
In particular the merging of track segments and legs of the same track is important.
Figure~\ref{fig:ex1} shows an example of a pp collision.
Some noisy TPC pads are clearly visible as stripes in the time direction.
Such noise clusters are actually dominant in single pp collisions, but are only a small fraction of the clusters in Pb--Pb collisions.
The tracker will find the different legs of looping tracks individually and then merge them into one track.
The removal of clusters of secondary legs works only if the merging works.
The figure shows multiple looping tracks, where the cluster removal worked only for the looping track in the lower part due to the merge failing for the other looping tracks.
This demonstrates the importance of the merging, and also the need to improve it.
On average around three instances of each low-$p_{\mathrm{T}}$ looping track are reconstructed, which means three legs are stored rather than one.
Numerical failures in the track fit during the synchronous phase (e.\,g.~through fake cluster attachment) must also be avoided.

\begin{figure}[!t]
  \centering
  \includegraphics[width=0.6\textwidth]{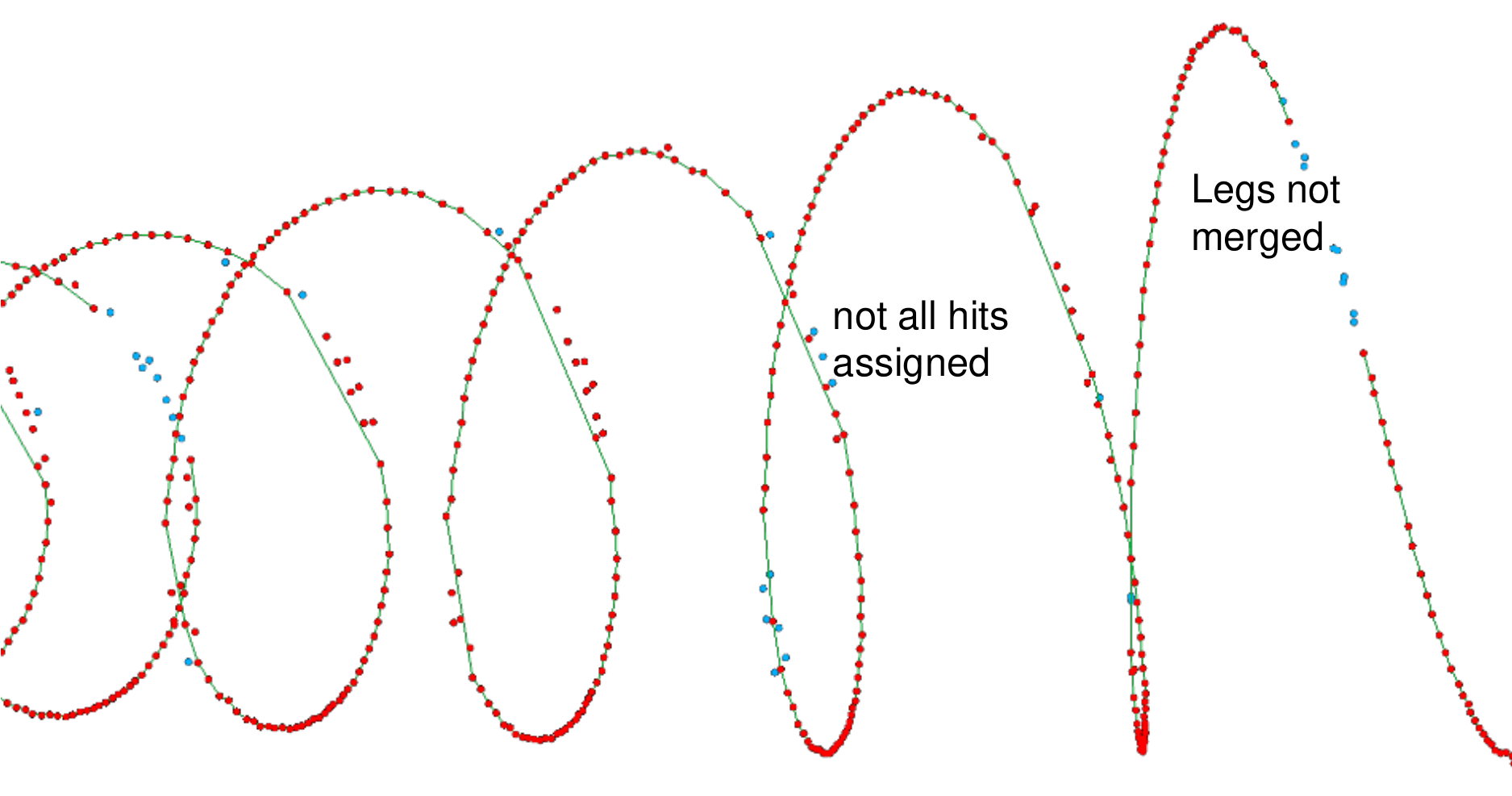}
  \caption{Detailed example of low-$p_{\mathrm{T}}$ tracking at the hand of a single low-$p_{\mathrm{T}}$ track.}
  \label{fig:ex2}
\end{figure}

Figure~\ref{fig:ex2} illustrates in more detail the challenges posed by a looping track with $p_{\mathrm{T}} < 50$ MeV/$c$.
Due to the low momentum, all attached and adjacent clusters are removed.
It also demonstrates that the merging failed at two places, yielding three track segments, and in some places the interpolation between the legs did not succeed to mark all clusters belonging to the track as adjacent.
The rejection ratio, in particular for strategy A, suffers accordingly.

\begin{figure}[!b]
  \centering
  \includegraphics[width=1.0\textwidth]{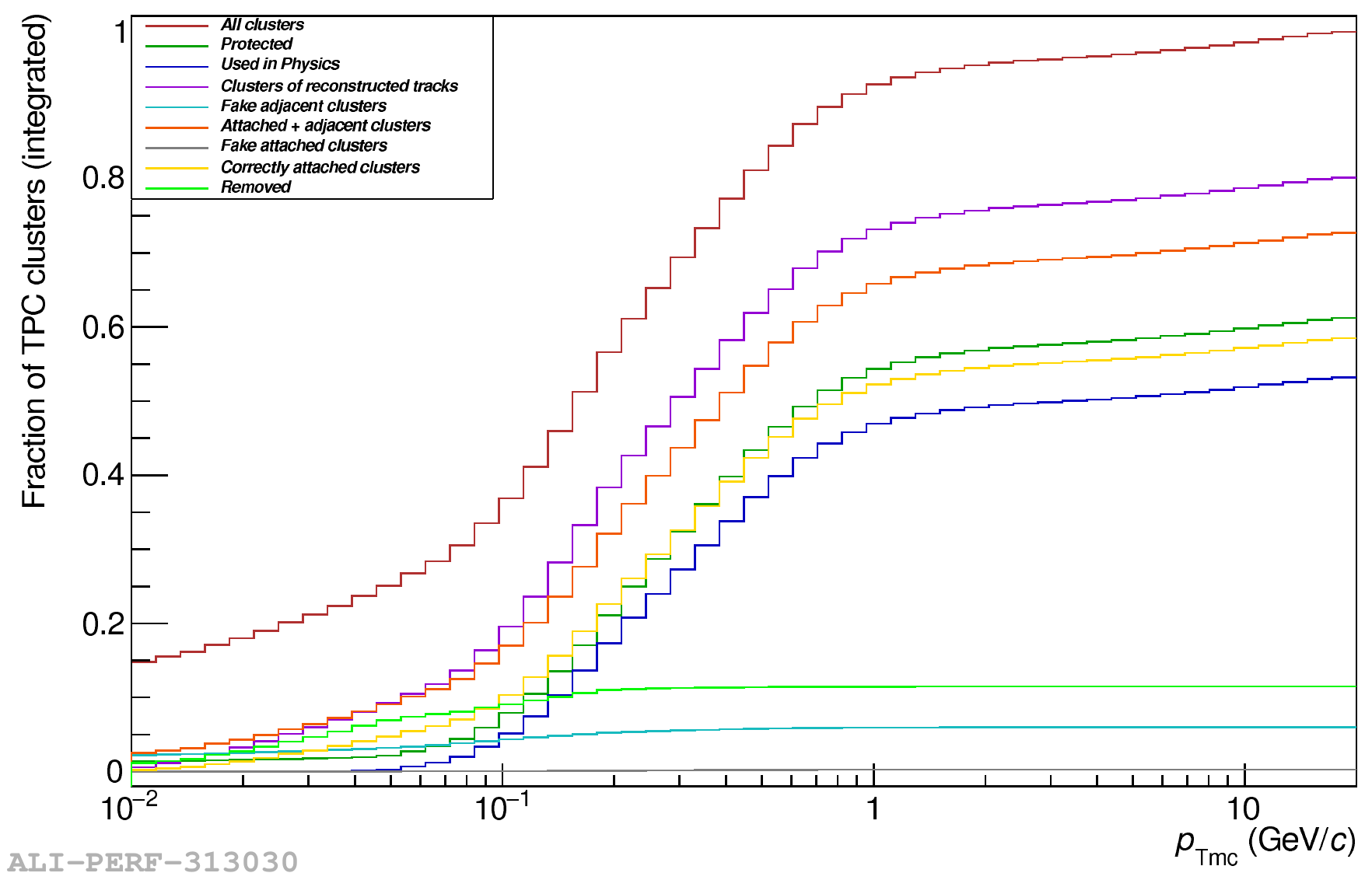}
  \caption{Integrated cluster attachment and rejection distribution of the ALICE TPC with the current development version of the Run~3 reconstruction software.}
  \label{fig:clusters}
\end{figure}

Figure~\ref{fig:clusters} gives an overview of the integrated cluster attachment and adjacent ratios in simulated Pb--Pb collisions, i.\,e.~what fraction of clusters below a $p_{\mathrm{T}}$ is marked as either attached or adjacent to a track and whether this assignment is correct or fake, versus the $p_{\mathrm{T}}$.
The curve for all clusters starts at alightly below 15\% at $p_{\mathrm{T}} = 10$ MeV/$c$, which shows that almost 15\% of the clusters stem from very low-$p_{\mathrm{T}}$ tracks, which are not accessible by tracking, and from noise.
The fake-attached clusters curve is nearly zero, demonstrating the quality of the tracking, which does not include fake hits in the fit.
However, the fake-adjacent clusters curve has a significant increase below 200 MeV/$c$.
This is unavoidable due to the presence of random high-$p_{\mathrm{T}}$ tracks in the vicinity of the clusters, which the cluster is marked as adjacent to.
This will be improved to some extent by replacing the fixed tube of 1.5 cm radius by a dynamic criterion taking into account the $\chi^2$.
The removable hits for strategy A are the difference between the orange line for attached + adjacent clusters and the dark green line for protected clusters (via tracking), and the offset of the all clusters line (via a specialized algorithm).
The purple line represents clusters of reconstructed tracks, irrespective of whether the cluster was attached to the track or not.
In other words, it assumes perfect cluster association for reconstructed tracks.
The difference between the clusters of reconstructed tracks and the attached + adjacent clusters represents the margin that we could gain with ideal cluster association in the tracking.

In numbers, 62.5\% of the clusters are protected with the current software, thus strategy B could remove 37.5\% of the hits today.
Improving the tube radius based on the $\chi^2$ could reduce the fake protected hits by 8\%.
In the current software state, strategy A can reject 12.5\% of the hits.
Perfect track merging would yield another 10\%,  and an ideal identification of junk below $p_{\mathrm{T}}$ = 10 MeV/$c$ could contribute an additional 13.5\% rejection.
Ideally, with perfect track merging, zero fake-protected hits, and perfect low-$p_{\mathrm{T}}$ junk identification, strategy A could remove 39.1\% and strategy B could remove 52.5\%.

\begin{figure}[!htb]
  \centering
  \includegraphics[width=0.5\textwidth]{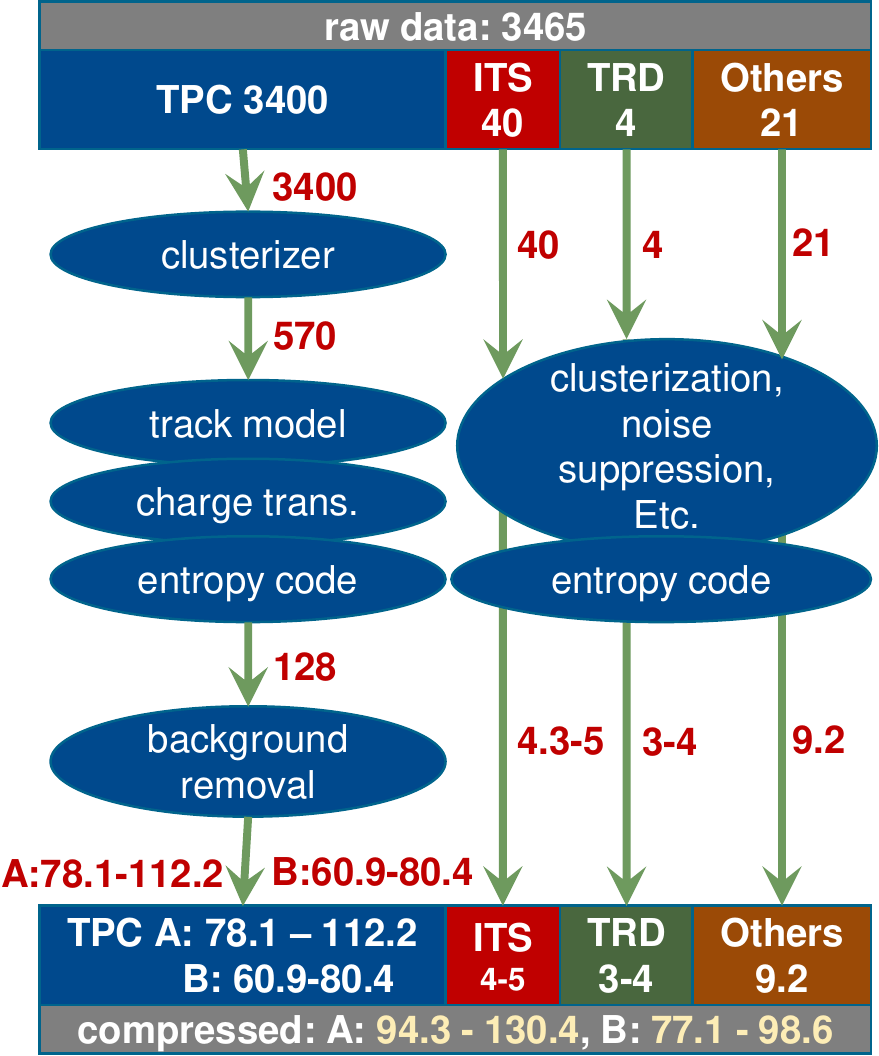}
  \caption{Overview of ALICE Run 3 peak data rates and compression during 50 kHz minimum-bias Pb--Pb data taking (1 GB = $10^9$ bytes).}
  \label{fig:rates}
\end{figure}

Figure~\ref{fig:rates} gives an overview of the data rates.
Here the TPC input data rates (and hence the total compression factor) must not be compared to that of Run 2 or to the TDR (Technical Design Report) since the data format has changed.
The 3400 GB/s correspond to raw ADC values in contrast to zero-suppressed data.
Also the 570 GB/s are to some extent arbitrary, as the output format of the clusterizer is not dense but contains padding.
This is a deliberate choice in order to simplify the processing.
The dense format which would correspond to step two of the compression explained in Sec.~\ref{sec:compression} yields 285 GB/s.
Accordingly, the ANS entropy compression achieves a compression factor of 2.2 yielding a rate of 128 GB/s.

\section{Conclusions}

The development of the tracking and the data compression for the ALICE Upgrade after LS2 are progressing well.
The mplementation of GPU alorithms for the baseline scenario is nearly finished.
The GPU-based TPC track finding shows a speedup of around 40 times compared to a single CPU core, while the optimization for the other GPU reconstruction steps is ongoing.
In parallel, we attempt to port more steps to the GPU for the full scenario.

While the other detectors are negligible with respect to the raw data rate, they become significant after taking into account the TPC compression.
Consequently, compression is implemented for all detectors as described in the TDR~\cite{bib:o2tdr}.
With the progress in the ANS entropy encoding for the TPC, we envision to use this as general entropy encoding step.
In addition, other detectors may run other data reduction steps beforehand just like the TPC.
The current projections for the output data rate with TPC rejection strategy B are already in agreement with the design goals from the TDR, while work for the more conservative strategy A is ongoing.






\begin{thebibliography}{99}

\bibitem{bib:alice}
{ALICE Collaboration},
J.~Inst. {\bf 3} S08002 (2008)

\bibitem{bib:aliceupgrade}
{ALICE Collaboration},
``{Upgrade of the ALICE Experiment: Letter of Intent}'',
CERN-LHCC-2012-012 (2012)

\bibitem{bib:tpcrun3tdr}
{ALICE Collaboration},
``Technical Design Report for the Upgrade of the ALICE Time Projection Chamber'',
CERN-LHCC-2013-020 (2013)

\bibitem{bib:o2tdr}
{ALICE Collaboration},
``Technical Design Report for the Upgrade of the Online-Offline Computing System'',
CERN-LHCC-2015-006, ALICE-TDR-019 (2015)

\bibitem{bib:hltpaper}
{ALICE Collaboration},
Real-time data processing in the ALICE High Level Trigger at the LHC,
submitted to Computer physics communications,
arXiv:1812:08036

\bibitem{bib:lhcp2017}
D.~Rohr for the {ALICE} Collaboration,
``{Tracking performance in high multiplicities environment at ALICE}'',
5th Large Hadron Collider Physics Conference (2017), arXiv:1709.00618

\bibitem{bib:itsgpu}
M.~Puccio for the {ALICE} Collaboration,
Proceedings of Science {\bf 287} 043 (2017)

\bibitem{bib:ctd2018}
D.~Rohr {\it et al.},
``Track Reconstruction in the ALICE TPC using GPUs for LHC Run 3'',
4th International Workshop Connecting the Dots (2018)
arXiv:1811.11481

\bibitem{bib:chep2018}
D.~Rohr {\it et al.},
``GPU-based Online Track Reconstruction for the ALICE TPC in Run~3 with Continuous Read-Out'',
CHEP (2018)
arXiv:1905:05515

\bibitem{bib:generic}
D.~Rohr {\it et al.},
``Portable and Vendor-Independent Low-Level Programming and Performance Benchmarking for Graphics Cards and Processors'',
IEEE HPCC Workshops, pp. 1-8 (2017)

\bibitem{bib:cnna}
D.~Rohr,
``ALICE TPC Online Tracker on GPUs for Heavy-Ion Events''
Proceedings of 13th International Workshop on Cellular Nanoscale Networks and their Applications, pp. 1-6 (2012)

\end{thebibliography}
\end{document}